# POWER LAWS IN BIOLOGICAL NETWORKS

Eivind Almaas[1] and Albert-László Barabási[2]

Department of Physics, University of Notre Dame, Notre Dame, IN  46556

[1]Almaas.1@nd.edu
[2]alb@nd.edu

## Abstract

The rapidly developing theory of complex networks indicates that real networks are not random, but have a highly robust large-scale architecture, governed by strict organizational principles. Here, we focus on the properties of biological networks, discussing their scale-free and hierarchical features. We illustrate the major network characteristics using examples from the metabolic network of the bacterium Escherichia coli. We also discuss the principles of network utilization, acknowledging that the interactions in a real network have unequal strengths. We study the interplay between topology and reaction fluxes provided by flux-balance analysis. We find that the cellular utilization of the metabolic network is both globally and locally highly inhomogeneous, dominated by "hot-spots", representing connected high-flux pathways.

**Introduction**

The tremendous progress in the natural sciences we witnessed in the last century was based on the reductionist approach, allowing us to predict the behavior of a system from the understanding of its (often identical) elementary constituents and their individual interactions. However, our ability to understand simple fundamental laws governing individual "building blocks" is a far cry from being able to predict the overall behavior of a complex system (Anderson, 1972). Additionally, the building blocks of most complex systems, and hence the nature of their interactions, vary dramatically, rendering the traditional approaches obsolete. During the last few years, network approaches have shown great promise as a new tool to analyze and understand complex systems (Strogatz, 2001; Albert, 2002; Dorogovtsev, 2003; Bornholdt, 2003). For example, technological information systems like the internet and the world-wide web are naturally modeled as networks, where the nodes are routers (Faloutsos, 1999; Vázquez, 2002) or web-pages (Albert, 1999; Lawrence, 1999; Broder, 2000) and the links are physical wires or URL's respectively. The analysis of societies also lends itself naturally to a network description, with people as nodes and the connections between the nodes as friendships (Milgram, 1967), collaborations (Kochen, 1989; Wasserman, 1994), sexual contacts (Liljeros, 2001) or co-authorship of scientific papers (Redner, 1998; Newman, 2001) to name a few possibilities. It seems that the closer we look at the world surrounding us, the more we realize that we are hopelessly entangled in myriads of interacting webs, and to describe them we need to understand the architecture of the various networks nature and technology offers us.

In biology, networks appear in many disparate systems, ranging from food webs in ecology to biochemical interactions in molecular biology. In particular in the cell the variety of interactions between genes, proteins and metabolites are well captured by networks. During the last decade, genomics has unleashed a downright flood of molecular interaction data. The nascent field of transcriptomics and proteomics have followed suit with analysis of protein levels under various conditions and genome wide analysis of gene expression at the mRNA level (Pandey, 2000; Caron, 2001; Burge, 2001). Thus, protein-protein interaction maps have been generated for a variety of organisms including viruses (Flajolet, 2000), prokaryotes like *H. pylori* (Rain, 2001) and eukaryotes like *S. cerevisiae* (Ito, 2000; Ito 2001; Schwikowski 2000; Uetz 2000; Gavin 2002, Ho, 2002, Jeong, 2001) and *C. elegans* (Walhout, 2000). In this chapter we will discuss recent results and developments in the study and characterization of naturally occurring networks, with focus on cellular ones.

**Power laws in network topology**

The complex network representation of different systems as networks has revealed surprising similarities, many of which are intimately tied to power laws. The simplest network measure is the average number of nearest neighbors of a node, or the average degree $\langle k \rangle$. However, this is a rather crude property, and to gain further insight into the topological organization of real networks, we need to determine the variation in the nearest neighbors, given by the degree distribution $P(k)$. For a surprisingly large number of networks, this degree distribution is best characterized by the power law functional form (Barabási, 1999) (Fig.1a),

$$P(k) \sim k^{-\alpha} . \qquad (1)$$

Important examples include the metabolic network of 43 organisms (Jeong, 2000), the protein interaction network of *S. cerevisiae* (Jeong, 2001) and various food webs (Montoya, 2002). If the degree distribution instead was single-peaked (e.g. Poisson or Gaussian) as in Fig. 1b, the majority of the nodes would be well described by the average degree, and hence the notion of a "typical" node. In contrast for networks with a power-law degree distribution, the majority of the nodes have only one or two neighbors while coexisting with many nodes with hundreds and some even with thousands of neighbors. For these networks there exists no typical node, and they are therefore often referred to as "scale-free".

The clustering of a node, the degree to which the neighborhood of a node resembles a complete subgraph, is another measure which sheds light on the structural organization of a network (Watts, 1998). For a node $i$ with degree $k_i$ the clustering is defined as

$$C_i = \frac{2n_i}{k_i(k_i - 1)}, \qquad (2)$$

representing the ratio of the number of actual connections between the neighbors of node $i$ to the number of possible connections. For a node which is part of a fully interlinked cluster $C_i = 1$, while $C_i = 0$ for a node which acts as a bridge between different clusters. Accordingly, the overall clustering coefficient of a network with $N$ nodes is given by $\langle C \rangle = \sum C_i / N$, and represents a measure of a network's potential modularity. By studying the clustering of nodes with a given degree $k$, information about the actual modular organization of a network can be gleaned (Ravasz 2002; Ravasz 2003; Dorogovtsev, 2002; Vázquez, 2002): For all metabolic networks available, this behaves like the power law

$$C(k) \sim k^{-\delta}, \qquad (3)$$

suggesting the existence of a hierarchy of nodes with different degrees of modularity (as measured by the clustering coefficient) overlapping in an iterative manner (Ravasz, 2002). In Fig. 2, we show the degree distribution (Fig. 2a) and the clustering as function of $k$ (Fig. 2b) for the bacterium *Escherichia coli*. They both clearly adhere to a power-law behavior, suggesting that biological networks are both scale-free and hierarchical. Panel 2c is a three dimensional representation of a cleaned up version of the metabolic network (Ravasz, 2002), demonstrating that modules are not clearly separated. Furthermore, the likelihood that a node appears in the shortest paths between other nodes on the network, the so-called betweenness-centrality $g$ (Freeman, 1977; Girvan, 2002), is also characterized by a power law distribution following $P(g) \sim g^{-\beta}$ for both biological and non-biological networks (Goh, 2002b), suggesting that a few nodes act as bridges or linkers between the different parts of the network . In summary, we have seen strong evidence that biological networks are both scale-free (Jeong, 2000; Jeong, 2001) and hierarchical (Ravasz, 2002).

**Network models**

An important question now arises – we can characterize networks using the above mentioned quantities, but why is the power law behavior so pervasive? Several models building on very different principles are able to explain these observed features.

*Random network models*

While graph theory initially focused on regular graphs, since the 1950's large networks with no apparent design principles were described as random graphs (Bollobas, 1985), proposed as the simplest and most straightforward realization of a complex network. According to the Erdos-Renyi (ER) model of random networks (Erdos, 1960), we start with $N$ nodes and connect every pair of nodes with probability $p$, creating a graph with approximately $pN(N-1)/2$ randomly distributed edges (Fig. 3a,d). For this model the degrees follow a Poisson distribution (Fig. 4a), and as a consequence, the average degree $\langle k \rangle$ of the network describes the typical node. Furthermore, for this "democratic" network model, the clustering is independent of the node degree $k$ (Fig. 4d). As we have just seen in Fig. 2, the ER model does not capture the properties of biological networks.

*Scale-free network model*

In the network model of Barabási and Albert (BA), two crucial mechanisms, which both are absent from the classical random network model, are responsible for the emergence of a power-law degree distribution (Barabási, 1999). First, networks grow through the addition of new nodes linking to nodes already present in the system. Second, there is a higher probability to link to a node with a large number of connections in most real networks, a property called preferential attachment. These two principles are implemented as follows: starting from a small core graph consisting of $m_0$ nodes, a new node with $m$ links is added at each time step and connected to the already existing nodes (Fig. 3b,e). Each of the $m$ new links are then preferentially attached to a node $i$ (with $k_i$ neighbors) which is chosen according to the probability

$$\Pi_i = k_i / \sum_j k_j . \qquad (4)$$

The simultaneous combination of these two network growth rules gives rise to the observed power-law degree distribution (Fig. 4b). In panel 3b, we illustrate the growth process of the scale-free model by displaying a network at time $t$ (green links) and then at time $(t+1)$, when we have added a new node (red links) using the preferential attachment probability. Compared to random networks, the probability that a node is highly connected is statistically significant in scale-free networks. Consequently, many network properties are determined by a relatively small number of highly connected nodes, often called "hubs". To make the effect of the hubs on the network structure visible, we have colored the five nodes with largest degrees red in Fig. 3d and 3e and their nearest neighbors green. While in the ER network only 27% of the nodes are reached by the five most connected ones, we reach more than 60% of the nodes in the scale-free network, demonstrating the key role played by the hubs. Another consequence of the hub's dominance of the network topology is that scale-free networks are highly tolerant of random failures (perturbations) while being extremely sensitive to targeted attacks (Albert, 2000). Comparing the properties of the BA network model with those of the ER model, we note that the clustering of the BA network is larger, however $C(k)$ is approximately constant (Fig. 4e), indicating the absence of a hierarchical structure.

*Hierarchical network model*

Many real networks are expected to be fundamentally modular, meaning that the network can be seamlessly partitioned into a collection of modules where each module performs an identifiable task, separable from the function(s) of other modules (Hartwell, 1999; Lauffenburger, 2000; Rao, 2001; Holter, 2001; Hasty, 2001; Shen-Orr, 2001). Therefore, we must reconcile the scale-free property with potential modularity. In order to account for the modularity as reflected in the power-law behavior of $C(k)$ (Fig. 2b) and a simultaneous scale-free degree distribution (Fig. 2a), we have to assume that clusters combine in an iterative manner, generating a hierarchical network (Ravasz, 2002; Vázquez, 2002). Such a network emerges from a repeated duplication and integration process of clustered nodes (Ravasz, 2002), which in principle can be repeated indefinitely. This process is depicted in panel 3c, where we start from a small cluster of four densely linked nodes (blue). We next generate three replicas of this hypothetical initial module (green) and connect the three external nodes of the replicated clusters to the central node of the old cluster, thus obtaining a large 16-node module. Subsequently, we again generate three replicas of this 16-node module (red), and connect the 16 peripheral nodes to the central node of the old module, obtaining a new module of 64 nodes. This hierarchical network model seamlessly integrates a scale-free topology with an inherent modular structure by generating a network that has a power law degree distribution (Fig. 4c) with degree exponent $\gamma = 1 + \ln 4 / \ln 3 \approx 2.26$ and a clustering coefficient $C(k)$ which proves to be dependent on $k^{-1}$ (Fig. 4f). However, note that modularity does not imply clear-cut sub-networks linked in well-defined ways (Ravasz, 2002; Holme, 2003). In fact, the boundaries of modules are often blurred (see Fig. 3f), bridged by highly connected nodes which interconnect modules.

**Power laws in network utilization**

Despite their successes, purely topologic approaches have important intrinsic limitations. For example, the activity of the various metabolic reactions or regulatory interactions differs widely, some being highly active under most growth conditions while others are switched on only for some rare environmental circumstances. Therefore, an ultimate description of cellular networks requires us to consider the intensity (i.e., strength), the direction (when applicable) and the temporal aspects of the interactions. While so far we know little about the temporal aspects of the various cellular interactions, recent results have shed light on how the strength of the interactions is organized in metabolic and genetic-regulatory networks (Almaas, 2004).

In metabolic networks the flux of a given metabolic reaction, representing the amount of substrate being converted to a product within unit time, offers the best measure of interaction strength. Recent metabolic flux-balance approaches (FBA) (Edwards, 2000; Edwards, 2001; Ibarra, 2002; Edwards, 2002; Segre, 2002) that allow us to calculate the flux for each reaction, have significantly improved our ability to generate quantitative predictions on the relative importance of the various reactions, leading to experimentally testable hypotheses. Starting from a stoichiometric matrix of the K12 MG1655 strain of *E. coli*, containing 537 metabolites and 739 reactions (Edwards, 2000; Edwards, 2001; Ibarra, 2002; Edwards, 2002), the steady state concentrations of all metabolites satisfy

$$\frac{d}{dt}[A_i] = \sum_j S_{ij} v_j = 0, \quad (5)$$

where $S_{ij}$ is the stoichiometric coefficient of metabolite $A_i$ in reaction $j$ and $v_j$ is the flux of reaction $j$. We use the convention that if metabolite $A_i$ is a substrate (product) in reaction $j$, $S_{ij} < 0$ ($S_{ij} > 0$), and we constrain all fluxes to be positive by dividing each reversible reaction into two "forward" reactions with positive fluxes. Any vector of positive fluxes $\{v_j\}$ which satisfies Eq. (5) corresponds to a state of the metabolic network, and hence, a potential state of operation of the cell.

Assuming that cellular metabolism is in a steady state and optimized for the maximal growth rate (Edwards, 2001; Ibarra, 2002), FBA allows us to calculate the flux for each reaction using linear optimization, providing a measure of each reaction's relative activity (Almaas, 2004). A striking feature of the flux distribution of *E. coli* is its overall inhomogeneity: reactions with fluxes spanning several orders of magnitude coexist under the same conditions (Fig. 4a). This is captured by the flux distribution for *E. coli,* which follows (the by now familiar) power law where the probability that a reaction has flux $v$ is given by $P(v) \sim (v + v_0)^{-\alpha}$. The flux exponent is predicted to be $\alpha = 1.5$ by FBA methods (Almaas, 2004). In a recent experiment (Emmerling, 2002) the strength of the various fluxes of the central metabolism was measured, revealing (Almaas, 2004) the power-law flux dependence $P(v) \sim v^{-\alpha}$ with $\alpha \cong 1$ (Fig. 4b). This power law behavior indicates that the vast majority of reactions have quite small fluxes, while coexisting with a few reactions with extremely large flux values.

The observed flux distribution is compatible with two quite different potential *local* flux structures (Almaas, 2004). A homogeneous local organization would imply that all reactions producing (consuming) a given metabolite have comparable fluxes. On the other hand, a more delocalized "hot backbone" is expected if the local flux organization is heterogeneous, such that each metabolite has a dominant source (consuming) reaction. To distinguish between these two scenarios for each metabolite $i$ produced (consumed) by $k$ reactions, we define the measure (Barthelemy, 2003; Derrida, 1987)

$$Y(k,i) = \sum_{j=1}^{k} \left( \frac{\hat{v}_{ij}}{\sum_{l=1}^{k} \hat{v}_{il}} \right)^2 , \qquad (6)$$

where $\hat{v}_{ij}$ is the mass carried by reaction $j$ which produces (consumes) metabolite $i$. If all reactions producing (consuming) metabolite $i$ have comparable $\hat{v}_{ij}$ values, $Y(k,i)$ scales as $1/k$. If, however, a single reaction's activity dominates Eq. (6), we expect $Y(k,i) \sim 1$, i.e., $Y(k,i)$ is independent of $k$. For the *E. coli* metabolism optimized for succinate and glutamate uptake (Fig. 5) we find that both the *in* and *out* degrees follow the power law $Y(k,i) \sim k^{-0.27}$, representing an intermediate behavior between the two extreme cases (Almaas, 2004). This indicates that the large-scale inhomogeneity observed in the overall flux distribution is increasingly valid at the level of the individual metabolites as well: the more reactions consume (produce) a given metabolite, the more likely it is that a single reaction carries the majority of the flux. This implies that the majority of the metabolic flux is carried along linear pathways – the metabolic high flux backbone (HFB) (Almaas, 2004).

A power law pattern is also observed when one investigates the strength of the various genetic regulatory interactions provided by microarray datasets. Assigning each pair of genes a correlation coefficient which captures the degree to which they are co-expressed, one finds that the distribution of these pair-wise correlation coefficients follows a power law (Kuznetsov, 2002; Farkas, 2003). That is, while the majority of gene pairs have only weak correlations, a few gene pairs display a significant correlation coefficient. These highly correlated pairs likely correspond to direct regulatory and protein interactions. This hypothesis is supported by the finding that the correlations are larger along the links of the protein interaction network and between proteins occurring in the same complex than for pairs of proteins that are not known to interact directly (Dezso, 2003; Grigoriev, 2001; Jansen, 2002; Ge, 2001).

Taken together, these results indicate that the biochemical activity in both the metabolic and genetic networks is dominated by several 'hot links' that represent a few high activity interactions embedded into a web of less active interactions. This attribute does not seem to be a unique feature of biological systems: hot links appear in a wide range of non-biological networks where the activity of the links follows a wide distribution (Goh, 2002a; deMenezes, 2004). The origin of this seemingly universal property is, again, likely rooted in the network topology. Indeed, it seems that the metabolic fluxes and the weights of the links in some non-biological system (Goh, 2002a; deMenezes, 2004) are uniquely determined by the scale-free nature of the network. A more general principle that could explain the correlation distribution data as well is currently lacking

**Conclusions**

Power laws are abundant in nature, affecting both the construction and the utilization of real networks. The power-law degree distribution has become the trademark of scale-free networks and can be explained by invoking the principles of network growth and preferential attachment. However, many biological networks are inherently modular, a fact which at first seems to be at odds with the properties of scale-free networks. However, these two concepts can co-exist in hierarchical scale-free networks. In the utilization of complex networks, most links represent disparate connection strengths or transportation thresholds. For the metabolic network of *E. coli* we can implement a flux-balance approach and calculate the distribution of link weights (fluxes), which (reflecting the scale-free network topology) displays a robust power-law, independent of exocellular perturbations. Furthermore, this global inhomogeneity in the link strengths is also present at the local level, resulting in a connected "hot-spot" backbone of the metabolism. Similar features are also observed in the strength of various genetic regulatory interactions. Despite the significant advances witnessed the last few years, network biology is still in its infancy, with future advances most notably expected from the development of theoretical tools, development of new interactive databases and increased insights into the interplay between biological function and topology.

E. Almaas and A.-L. Barabási    12

**References**

Albert, R. & Barabási, A.-L. (2002). Statistical mechanics of complex networks. *Rev. Mod. Phys.* 74, p47-97.

Albert, R., Jeong, H. & Barabási, A.-L (1999). Diameter of the World-Wide Web. *Nature*, 401, p130-1.

Albert, R., Jeong, H. & Barabási, A.-L. (2000). Attack and error tolerance of complex networks. *Nature*, 406, p378-82.

Almaas, E., Kovacs, B., Vicsek, T., Oltvai, Z.N. & Barabási, A.-L. (2004). Global organization of metabolic fluxes in the bacterium *Escherichia coli*. *Nature*, in press.

Anderson, P. W. (1972). More Is Different. *Science*, 177, p393-6.

Barabási, A.-L. & Albert, R., (1999). Emergence of scaling in random networks. *Science*, 286, p509-12.

Barthelemy, M., Gondran, B. & Guichard, E. (2003). Spatial structure of the Internet traffic. *Physica A*, 319, p633-42.

Bollobas, B. (1985). *Random Graphs*. Academic Press, London.

Bornholdt, S. & Schuster, H. G. (2003). *Handbook of graphs and networks: From the genome to the Internet*. Wiley-VCH, Berlin, Germany.

Broder, A., Kumar, R., Maghoul, F., Raghavan, P, Rajalopagan, S., Stata, R., Tomkins, A. & Wiener, J. (2000). Graph structure in the web. *Comput. Netw.*, 33, p309-20.

Burge, C. (2001). Chipping away at the transcriptome. *Nature Genet.*, 27, p232-4.

Caron, H., van Schaik, B., van der Mee, M., Baas, F., Riggins, G., van Sluis, P., Hermus, M.C., van Asperen, R., Boon, K., Voute, P.A., Heisterkamp, S., van Kampen, A. & Versteeg, R. (2001). The human transcriptome map: Clustering of highly expressed genes in chromosomal domains. *Science*, 291, p1289-92.

deMenezes, M.A. & Barabási, A.-L. (2004) Fluctuations in network dynamics. *Phys. Rev. Lett.*, in press.

Derrida, B. & Flyvbjerg, H. (1987). Statistical properties of randomly broken objects and of multivalley structures in disordered-systems. *J. Phys. A: Math. Gen.*, 20, p5273-88 (1987).

E. Almaas and A.-L. Barabási                                                                 13Dezso, Z., Oltvai, Z.N. & Barabási, A.-L. (2003) Bioinformatics analysis of experimentally determined protein complexes in the yeast, *Saccharomyces cerevisiae*. *Genome Res.*, 13, p2450-4.

Dorogovtsev, S.N., Goltsev, A.V. & Mendes, J.F.F. (2002). Pseudofractal scale-free web. *Phys. Rev. E*, 65, 066122.

Dorogovtsev, S.N. & Mendes, J.F.F. (2003) *Evolution of networks: From biological nets to the Internet and WWW*. Oxford University Press, Oxford.

Edwards, J. S., Ibarra, R. U. & Palsson, B. O. (2001). In silico predictions of Escherichia coli metabolic capabilities are consistent with experimental data. *Nat Biotechnol* **19**, p125-30.

Edwards, J. S. & Palsson, B. O. (2000). The Escherichia coli MG1655 in silico metabolic genotype: its definition, characteristics, and capabilities. *Proc Natl Acad Sci U S A* 97, p5528-33.

Edwards, J. S., Ramakrishna, R. & Palsson, B. O. (2002). Characterizing the metabolic phenotype: A phenotype phase plane analysis. *Biotechn. Bioeng.* 77, 27-36.

Emmerling, M., Dauner, M., Ponti, A., Fiaux, J., Hochuli, M., Szyperski, T., Wuthrich, K., Bailey, J.E. & Sauer, U. (2002). Metabolic flux responses to pyruvate kinase knockout in Escherichia coli. *J Bacteriol.*, 184, p152-64.

Erdos, P. & Renyi, A. (1960). On the evolution of random graphs. *Publ. Math. Inst. Hung. Acad. Sci.*, 5, p17-61.

Faloutsos, M., Faloutsos, P. & Faloutsos, C. (1999). On power-law relationships of the Internet topology. *Comput. Commun. Rev.*, 29, p251-62.

Farkas, I.J., Jeong, H., Vicsek, T., Barabási, A.-L. & Oltvai, Z.N. (2003). The topology of the transcription regulatory network in the yeast, *Saccharomyces cerevisiae*. *Physica A*, 318, p601-12.

Flajolet, M., Rotondo, G., Daviet, L., Bergametti, F., Inchauspe, G., Tiollais, P., Transy, C. & Legrain, P. (2000). A genomic approach to the hepatitis C virus. *Gene*, 242, p369-79.

Freeman, L. (1977). A set of measures of centrality based upon betweenness. *Sociometry*, 40, p35-41.

E. Almaas and A.-L. Barabási    14

Gavin, A.C., Bosche, M., Krause, R., Grandi, P., Marzioch, M., Bauer, A., Schultz, J., Rick, J.M., Michon, A.M., Cruciat, C.M., Remor, M., Hofert, C., Schelder, M., Brajenovic, M., Ruffner, H., Merino, A., Klein, K., Hudak, M., Dickson, D., Rudi, T., Gnau, V., Bauch, A., Bastuck, S., Huhse, B., Leutwein, C., Heurtier, M.A., Copley, R.R., Edelmann, A., Querfurth, E., Rybin, V., Drewes, G., Raida, M., Bouwmeester, T., Bork, P., Seraphin, B., Kuster, B., Neubauer, G. & Superti-Furga, G. (2002). Functional organization of the yeast proteome by systematic analysis of protein complexes. *Nature*, 415, p141-7.

Ge, H. Liu, Z., Church, G.M. & Vidal, M. (2001). Correlation between transcriptome and interactome mapping data from *Saccharomyces cerevisiae*. *Nature Genet.*, 29, p482-6.

Girvan, M. & Newman, M.E.J. (2002). Community structure in social and biological networks. *Proc. Natl. Acad. Sci.*, 99, p7821-26.

Goh, K.-I., Kahng, B. & Kim, D. (2002a). Fluctuation-driven dynamics of the internet topology. *Phys. Rev. Lett.*, 88, 108701.

Goh, K.-I., Oh, E., Jeong, H., Kahng, B. & Kim, D. (2002b). Classification of scale-free networks. *Proc. Natl. Acad. Sci.*, 99, p12583-88.

Grogoriev, A. (2001). A relationship between gene expression and protein interactions on the proteome scale: analysis of the bacteriophage T7 and yeast *Saccharomyces cerevisiae*. *Nucleic Acids Res.*, 29, p3513-9.

Hartwell, L.H., Hopfield, J.J., Leibler, S. & Murray, A.W. (1999). From molecular to modular cell biology. *Nature*, 402, C47-52.

Hasty, J., McMillen, D., Isaacs, F. & Collins, J.J. (2001). Computational studies of gene regulatory networks: In numero molecular biology. *Nature Rev. Genet.*, 2, p268-79.

Ho, Y., Gruhler, A., Heilbut, A., Bader, G.D., Moore, L., Adams, S.L., Millar, A., Taylor, P., Bennett, K., Boutilier, K., Yang, L.Y., Wolting, C., Donaldson, I., Schandorff, S., Shewnarane, J., Vo, M., Taggart, J., Goudreault, M., Muskat, B., Alfarano, C., Dewar, D., Lin, Z., Michalickova, K., Willems, A.R., Sassi, H., Nielsen, P.A., Rasmussen, K.J., Andersen, J.R., Johansen, L.E., Hansen, L.H., Jespersen, H., Podtelejnikov, A., Nielsen, E., Crawford, J., Poulsen, V., Sorensen, B.D., Matthiesen, J., Hendrickson, R.C., Gleeson, F., Pawson, T., Moran, M.F., Durocher, D., Mann,

M., Hogue, C.W.V., Figeys, D. & Tyers, M. (2002). Systematic identification of protein complexes in Saccharomyces cerevisiae by mass spectrometry. *Nature*, 415, p180-3.

Holme, P., Huss, M. & Jeong, H. (2003). Subnetwork hierarchies of biochemical pathways. *Bioinformatics*. 19, p532-9.

Holter, N.S., Maritan, A., Cieplak, M., Fedoroff, N.V. & Banavar, J.R. (2001). Dynamic modeling of gene expression data. *Proc. Natl. Acad. Sci.*, 98, p1693-8.

Ibarra, R. U., Edwards, J. S. & Palsson, B. O. (2002). Escherichia coli K-12 undergoes adaptive evolution to achieve in silico predicted optimal growth. *Nature* 420, p186-9.

Ito, T., Chiba, T., Ozawa, R., Yoshida, M., Hattori, M. & Sakaki, Y. (2001). A comprehensive two-hybrid analysis to explore the yeast protein interactome. *Proc. Natl. Acad. Sci.*, 98, p4569-74.

Ito, T., Tashiro, K., Muta, S., Ozawa, R., Chiba, T., Nishizawa, M., Yamamoto, K., Kuhara, S. & Sakaki, Y. (2000). Towards a protein-protein interaction map of the budding yeast: A comprehensive system to examine two-hybrid interactions in all possible combinations between the yeast proteins. *Proc. Natl. Acad. Sci.*, 97, p1143-47.

Jansen, R., Greenbaum, D. & Gerstein, M. (2002). Relating whole-genome expression data with protein-protein interactions. *Genome Res.*, 12, p37-46.

Jeong, H., Mason, S.P., Barabási, A.-L. & Oltvai, Z.N. (2001). Lethality and centrality in protein networks. *Nature*, 411, p41-2.

Jeong, H., Tombor, B., Albert, R., Oltvai, Z.N. & Barabási, A.-L. (2000). The large-scale organization of metabolic networks. *Nature*, 407, p651-4.

Kochen, M. (ed.) (1989). *The small-world*. Ablex, Norwood, N.J.

Kutznetsov, V.A., Knott, G.D. & Bonner, R.F. (2002). General statistics of stochastic processes of gene expression in eukaryotic cells. *Genetics*, 161, p1321-32.

Lauffenburger, D. (2000). Cell signaling pathways as control modules: Complexity for simplicity. *Proc. Natl. Acad. Sci.*, 97, p5031-33.

Lawrence, S. & Giles, C. L. (1999). Accessibility of information on the web. *Nature*, 400, p107-9.

Liljeros, F., Edling, C.R., Amaral, L.A.N., Stanley, H.E. Aberg, Y. (2001). The web of human sexual contacts. *Nature*, 411, p907-8.

McGraith, S., Holtzman, T., Moss, B. & Fields, S. (2000). Genome-wide analysis of vaccinia virus protein-protein interactions. *Proc. Natl. Acad. Sci.*, 97, p4879-84.

Milgram, S. (1967). The small-world problem. *Psychology Today*, 2, p60-7.

Montoya, J.M. & Sole, R.V. (2002). Small-world patterns in food webs. *J. Theor. Biol.*, 214, p405-12.

Newman, M.E.J. (2001). The structure of scientific collaboration networks. *Proc. Natl. Acad. Sci.*, 98, p404-9.

Pandey, A. & Mann, M. (2000). Proteomics to study genes and genomes. *Nature*, 405, p837-46.

Rain, J.-C., Selig, L., DeReuse, H., Battaglia, V., Reverdy, C., Simon, S., Lenzen, G., Petel, F., Wojcik, J., Schächter, V., Chemama, Y., Labigne, A. & Legrain, P. (2001). The protein-protein interaction map of *Helicobacter pylori*. *Nature*, 409, p211-15.

Rao, C.V. & Arkin, A.P. (2001). Control motifs for intracellular regulatory networks. *Annu. Rev. Biomed. Eng.*, 3, p391.

Ravasz, E., Somera, A.L., Mongru, D.A., Oltvai, Z.N. & Barabási, A.-L. (2002). Hierarchical organization of modularity in metabolic networks. *Science*, 297, p1551-5.

Ravasz, E. & Barabási, A.-L. (2003). Hierarchical organization in complex networks. *Phys. Rev. E*, 67, 026112.

Schwikowski, B., Uetz, P., & Fields, S. (2000). A network of protein-protein interactions in yeast. *Nature Biotechn.*, 18, p1257-61.

Segre, D., Vitkup, D. & Church, G. M. (2002). Analysis of optimality in natural and perturbed metabolic networks. *Proc. Natl. Acad. Sci.*, 99, p15112-7.

Shen-Orr, S.S., Milo, R., Mangan, S. & Alon, U. (2001). Network motifs in the transcriptional regulation network of *Escherichia coli*. *Nature Genet.*, 31, p64-8.

Strogatz, S.H. (2001). Exploring complex networks. *Nature*, 410, p268-76.

Uetz, P., Giot, L., Cagney, G., Mansfield, T., Judson, R., Knight, J., Lockshorn, D., Narayan, V., Srinivasan, M., Pochart, P., Qureshi-Emili, A., Li, Y., Godwin, B., Conover, D., Kalbfleisch, T., Vijayadamodar, G., Yang, M.J., Johnston, M., Fields, S.

& Rothberg, J.M. (2000). A comprehensive analysis of protein-protein interactions in Saccharomyces cerevisiae. *Nature*, 403, p623-27.

Vázquez, A., Pastor-Satorras, R. & Vespignani, A. (2002). Large-scale topological and dynamical properties of the Internet. *Phys. Rev. E*, 65, 066130.

Walhout, A., Sordella, R., Lu, X., Hartley, J., Temple, G., Brasch, M., Thierry-Mieg, N., & Vidal, M. (2000). Protein interaction mapping in *C. elegans* using proteins involved in vulva development. *Science*, 287, p116-22.

Wasserman, S. & Faust, K. (1994). *Social Network Analysis: Methods and Applications.* Cambridge University Press, Cambridge.

Watts, D.J. & Strogatz, S.H. (1998). Collective dynamics of small-world networks. *Nature*, 393, p440-2.

**FIGURE CAPTIONS**

**Figure 1.** Characterizing degree distributions. For the power-law degree distribution **(a)**, there exists no typical node, while for single peaked distributions **(b)**, most nodes are well represented by the average (typical) node with degree $\langle k \rangle$.

**Figure 2.** Properties of the metabolic network of *Escherichia coli*. **(a)** The degree distribution displays a power law in both the in- and the out degrees (Jeong, 2000). **(b)** The clustering coefficient varies with *k* as a power law. The solid line corresponds to $k^{-1}$. **(c)** Three dimensional representation of the reduced metabolic network (Ravasz, 2002).

**Figure 3.** Graphical representation of three network models: **(a)** and **(d)** The ER (random) model, **(b)** and **(e)** the BA (scale-free) model and **(c)** and **(f)** the hierarchical model. The random network model is constructed by starting from *N* nodes before the possible node-pairs are connected with probability *p*. Panel **(a)** shows a particular realization of the ER model with 10 nodes and connection probability $p = 0.2$. In panel **(b)** we show the scale-free model at time *t* (green links) and at time $(t+1)$ when we have added a new node (red links) using the preferential attachment probability (see Eq. (4)). Panel **(c)** demonstrates the iterative construction of a hierarchical network, starting from a fully connected cluster of four nodes (blue). This cluster is then copied three times (green) while connecting the peripheral nodes of the replicas to the central node of the starting cluster. By once more repeating this replication and connection process (red nodes), we end up with a 64-node scale-free hierarchical network. In panel **(d)** we display a larger version of the random network, and it is evident that most nodes have approximately the same number of links. For the scale-free model, **(e)** the network is clearly inhomogeneous: while the majority of nodes has one or two links, a few nodes have a large number of links. We emphasize this by coloring the five nodes with the highest number of links red and their first neighbors green. While in the random network only 27% of the nodes are reached by the five most connected nodes, we reach more than 60% of the nodes in the scale-free network, demonstrating the key role played by the hubs. Note that the networks in **(d)** and **(e)** consist of the same number of nodes and links. Panel **(f)** demonstrates that the standard clustering algorithms are not that successful in uncovering the modular structure of a scale-free hierarchical network.

**Figure 4.** Properties of the three network models. **(a)** The ER model sports a Poisson degree distribution *P(k)* (the probability that a randomly selected node has exactly *k* links) which is strongly peaked at the average degree $\langle k \rangle$ and decays exponentially for large *k*. The degree distributions for the scale-free **(b)** and the hierarchical **(c)** network models do not have a peak, they instead decay according to the power-law $P(k) \sim k^{-\gamma}$. The average clustering coefficient for nodes with exactly *k* neighbors, *C(k)*, is independent of *k* for both the ER **(d)** and the scale-free **(e)** network model. **(f)** In contrast, $C(k) \sim k^{-1}$ for the hierarchical network model (cf. Fig. 2).

**Figure 5.** Flux distribution for the metabolism of *E. coli*. **(a)** Flux distribution for optimized biomass production on succinate (black) and glutamate (red) rich uptake substrates. The solid line corresponds to the power law fit $P(v) \sim (v + v_0)^{-\alpha}$ with $v_0 = 0.0003$ and $\alpha = 1.5$. **(b)** The distribution of experimentally determined fluxes (see Emmerling (2002)) from the central metabolism of *E. coli* also displays power-law behavior with a best fit to $P(v) \sim v^{-\alpha}$ with $\alpha = 1$.

**Figure 6.** Characterizing the local inhomogeneity of the metabolic flux distribution. The measured *kY(k)* (see Eq. (6)) shown as function of *k* for incoming and outgoing reactions for fluxes calculated on both succinate and glutamate rich substrates, averaged over all metabolites, indicating $Y(k) \sim k^{-0.27}$, as the straight line in the figure has slope $\gamma = 0.73$. Inset: The non-zero mass flows $\hat{v}_{ij}$ producing (consuming) flavin adenine dinucleotide (FAD) on a glutamate rich substrate.

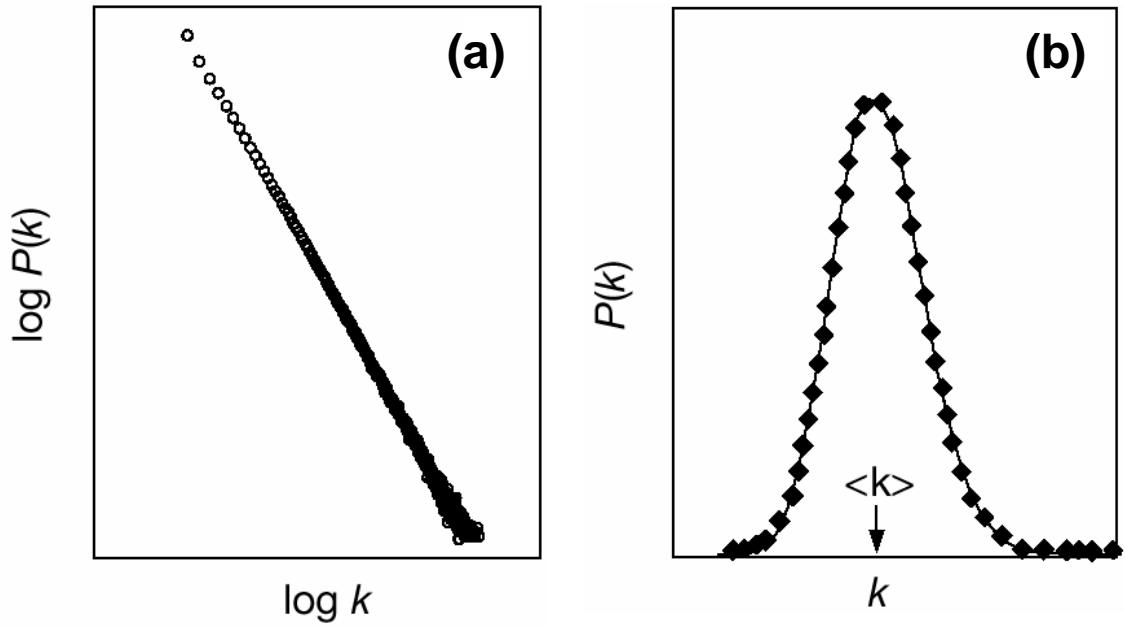

**Figure 1.**

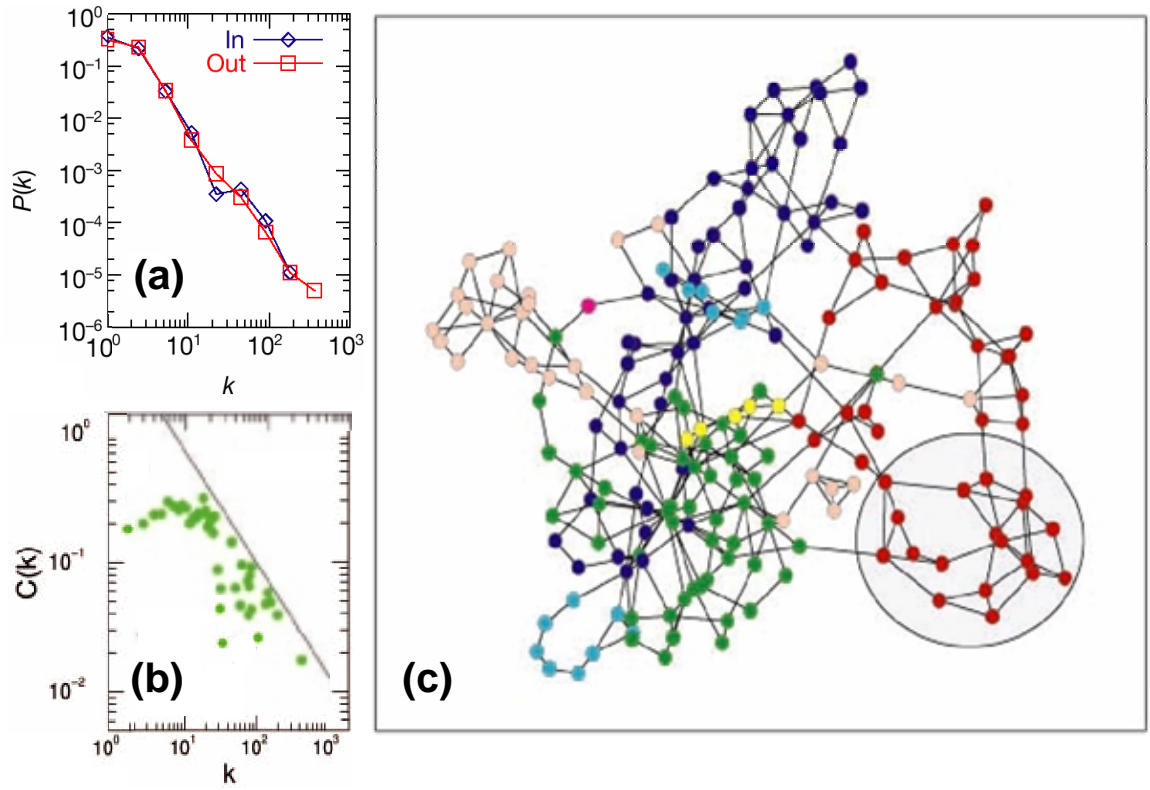

**Figure 2.**

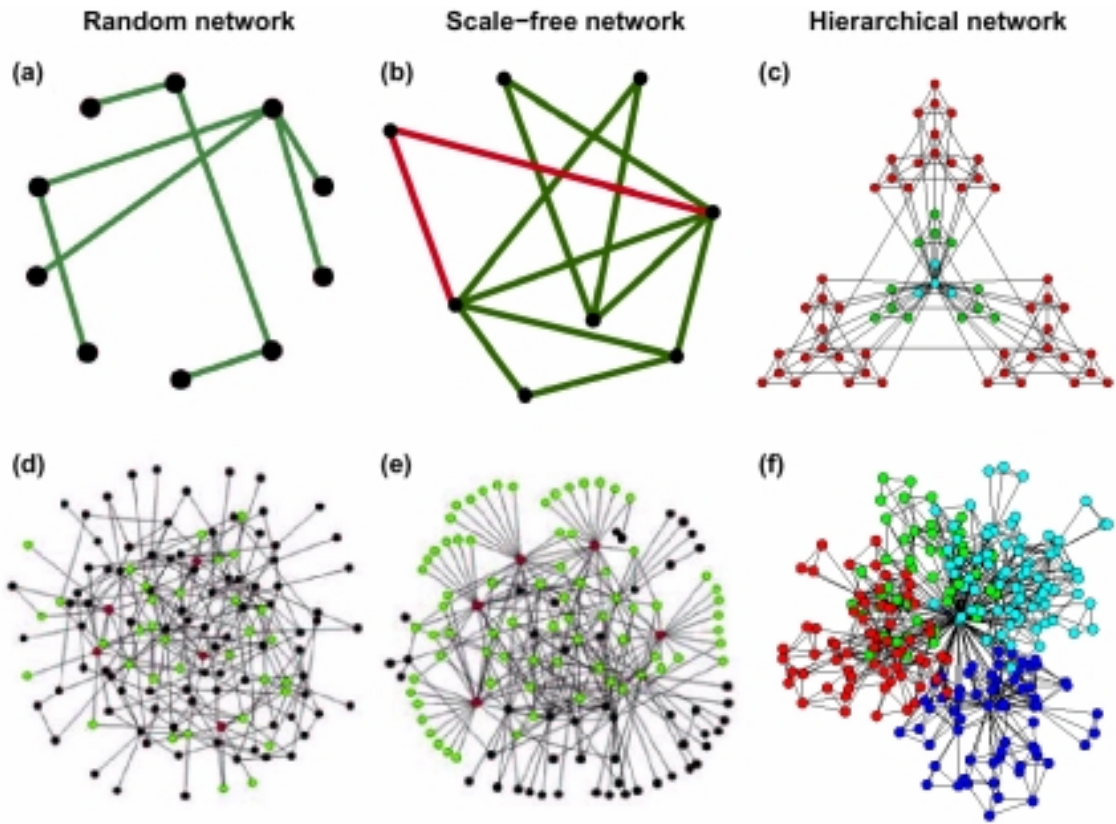

**Figure 3.**

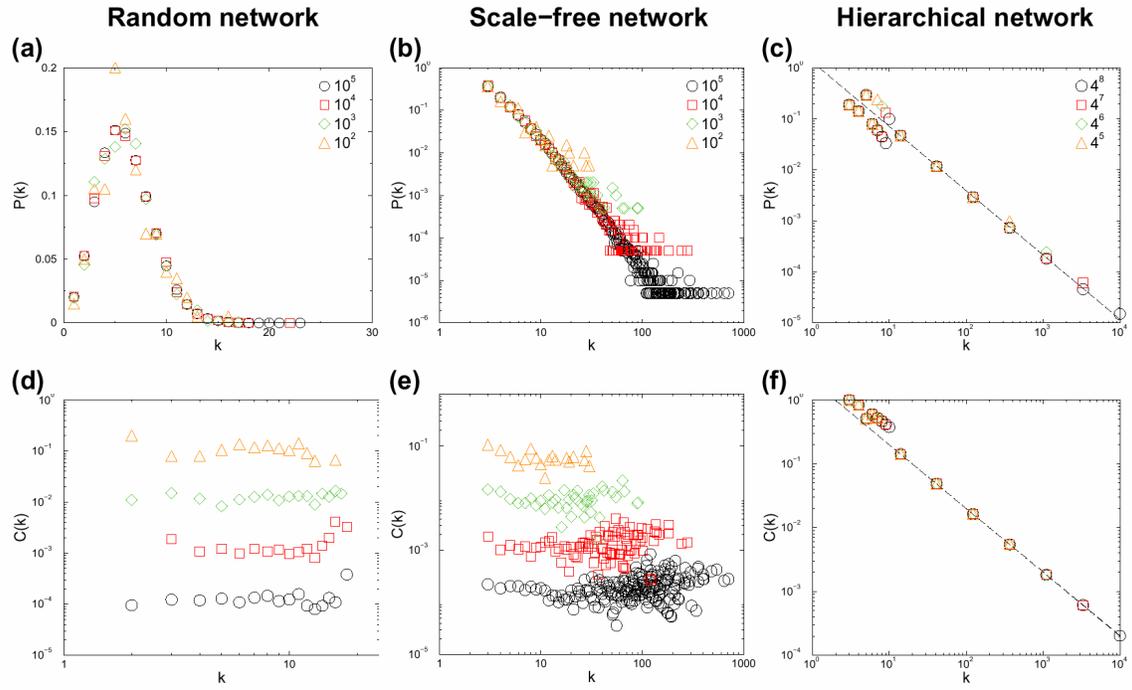

**Figure 4.**

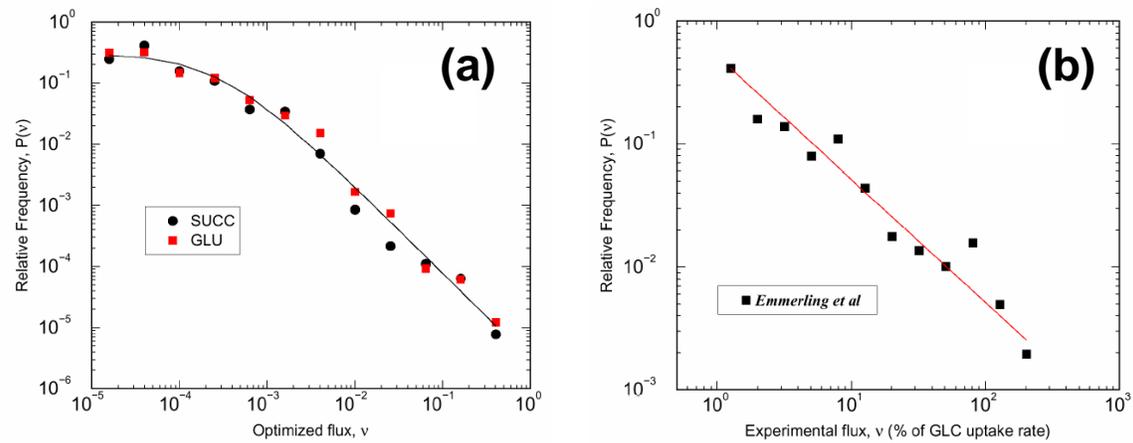

**Figure 5.**

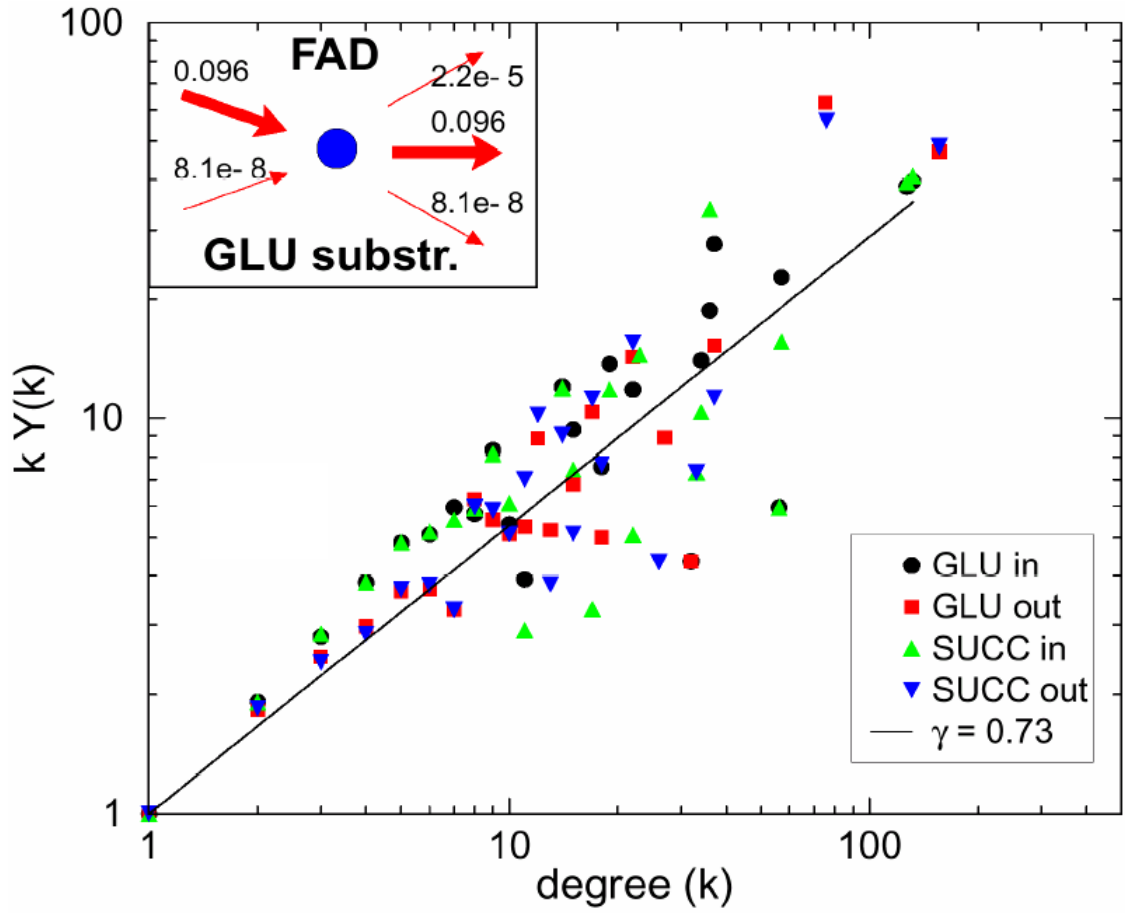

**Figure 6.**